\documentclass[aip,reprint,bibliography,superscriptaddress]{revtex4-1}
\usepackage{mathrsfs}
\usepackage{amsmath,gensymb}
\usepackage{amsfonts}
\usepackage{amssymb}
\usepackage{amsthm}
\usepackage{graphicx}
\usepackage{natbib}
\usepackage{xcolor}
\usepackage{hyperref}
\usepackage{bm}
\usepackage[caption=false]{subfig}
\usepackage{verbatim}

\begin{document}

\title{Vortex Dynamics Driven Transverse Transport in Superconducting Thin Films}
\author{Laxmipriya Nanda}
\affiliation{Centre for Nanoscience and Engineering, Indian Institute of Science, Bengaluru, Karnataka 560012, India} 
\author{Sohini Guin}
\affiliation{Centre for Nanoscience and Engineering, Indian Institute of Science, Bengaluru, Karnataka 560012, India} 
\author{Yasen Hou}
\affiliation{Plasma Science and Fusion Center, Massachusetts Institute of Technology, Cambridge 02139, USA} 
\author{Rajib Sarkar}
\affiliation{Centre for Nanoscience and Engineering, Indian Institute of Science, Bengaluru, Karnataka 560012, India} 
\author{Naresh Shyaga}
\affiliation{Centre for Nanoscience and Engineering, Indian Institute of Science, Bengaluru, Karnataka 560012, India} 
\author{Souvik Banerjee}
\affiliation{Jawaharlal Nehru Center for Advanced Scientific Research, Jakkur, Bengaluru}
\author{A. Sundaresan}
\affiliation{Jawaharlal Nehru Center for Advanced Scientific Research, Jakkur, Bengaluru}
\author{N. S. Vidhyadhiraja}
\affiliation{Jawaharlal Nehru Center for Advanced Scientific Research, Jakkur, Bengaluru}
\author{Jagadeesh S. Moodera}
\affiliation{Plasma Science and Fusion Center, Massachusetts Institute of Technology, Cambridge 02139, USA} 
\affiliation{Department of Physics, Massachusetts Institute of Technology, Cambridge 02139, USA} 
\author{Dhavala Suri}
\email{dsuri@iisc.ac.in}
\affiliation{Centre for Nanoscience and Engineering, Indian Institute of Science, Bengaluru, Karnataka 560012, India}

\begin{abstract}
The generation of transverse voltage in superconducting thin films carrying a longitudinal current has been recently related to demonstrate exotic, symmetry-breaking transport phenomena. To understand the underlying phenomenon we chose two distinctly different systems to investigate; a conventional (Nb) and a nonconventional (Ni/Bi bilayer) superconductor. A systematic study of this transverse response in these type-II superconducting thin films we aimed towards establishing its microscopic origin. Both longitudinal and transverse magnetotransport measurements were carried out with Hall-bar patterned devices as a function of temperature with magnetic field applied  out-of-plane. In both systems, a sharp, field antisymmetric transverse resistance emerges in the superconducting transition region, closely following the field derivative of the longitudinal resistance while it was vanishingly small or absent in the fully superconducting or normal states. Based on the vortex-flow model, the behavior could be quantitatively attributed as the transverse voltage arising due to the motion of Abrikosov vortices under a combined action of Magnus like force and viscous drag force, rather than from any exotic or intrinsic Hall-type mechanism. The close correspondence between a nominally conventional superconductor Nb and an unconventional superconductor heterostructure incorporating a ferromagnetic layer indicates that the vortex dynamics alone, independent of magnetic proximity effects or unconventional pairing, seems to be sufficient to account for the observed transverse signal.
Our observations while seems to provide a quantitative, material-independent model for interpreting transverse voltages in superconducting films, raises the concerns in attributing such signals to exotic physics without first ruling out conventional vortex-flow contributions.
\end{abstract}

\maketitle

Type-II superconductors respond to an applied magnetic field by admitting quantized units of magnetic flux, known as Abrikosov vortices, once the field exceeds the lower critical field $H_{c1}$~\cite{abrikosov1957magnetic, tinkham2004, brandt2002vortices}. In this mixed state, the interplay between an applied current and the resulting vortex lattice governs the electrical response of the superconductor: vortices set into motion by the Lorentz force dissipate energy and generate a measurable voltage, even though the material remains microscopically superconducting~\cite{bardeen1965, tinkham2004, huebener1994thermal, Brandt1995}. Understanding this vortex-mediated dissipation is therefore essential not only for characterizing the fundamental superconducting state but also for interpreting transport anomalies that have, in several recent material systems, been proposed as evidence of more exotic, symmetry-breaking physics\cite{van1995vortex,matsuda1995hall,davydova2024,nadeem2023}.

Transport measurements offer a direct and sensitive probe of the superconducting condensate and its excitations, particularly near the transition where fluctuations and vortex dynamics dominate~\cite{skocpol1975,blatter1994vortices,varlamov2018,tinkham2004,lang1995study}. In an ideal superconducting state, no Hall response is expected; however, in the mixed state close to $T_c$, finite transverse voltages can arise from vortex motion and quasiparticle backflow~\cite{Nozieres1966, kopnin1995}.

The symmetry and field dependence of such transverse resistance anomalies provide insight into the dimensionality of superconductivity and the nature of dissipation mechanisms~\cite{vinokur1993, wang1991anomalous}. The transverse voltage response of superconductors in the mixed state has been studied from two rather different perspectives. Segal, Karpovski, and Gerber approached the problem from a materials-inhomogeneity standpoint, showing that even slight, unintentional inhomogeneity in an otherwise conventional superconductor can generate spurious zero-field and even-in-field transverse voltage signals near the superconducting transition, attributable to a simple circuit model in which additional voltage components scale with the temperature and field derivatives of the longitudinal resistivity \cite{Segal2011}. This picture was recently extended by Sengupta \textit{et al.}, who found that current paths become highly nonuniform at the onset of the superconducting transition in conventional superconducting films, producing a finite transverse resistance from macroscopic electronic inhomogeneities unrelated to the films' structural properties, and establishing this as a ubiquitous phenomenon rather than a sample-specific artifact \cite{sengupta2025}. In contrast, Xu, Li, and Chien reported an anomalous transverse voltage in epitaxial $\beta$-Bi$_2$Pd thin films that is not of this extrinsic origin: the sign of the signal is independent of the applied magnetic field and is instead dictated by the broken inversion symmetry at the film interfaces, pointing to an intrinsic chirality tied to topological surface states \cite{Xu2022}. To shed light on the competing origin of transverse voltage observed near a superconducting transition resulting from extrinsic inhomogeneity versus intrinsic, symmetry-protected chirality, we undertook the current studies.

In this context, Ni/Bi and Nb are two archetypal materials of type-II superconductors with dynamics driven by vortices\cite{reimann2015,liu2020magnetic}. Early experiments reported superconductivity in Ni/Bi with $T_c \sim 4$~K\cite{Leclair_2005} and clear deviations from conventional $s$-wave behavior\cite{Gong_2015}, with point-contact Andreev reflection measurements suggesting spin-triplet $p$-wave pairing, possibly of the Anderson-Brinkman-Morel type~\cite{Gong_2015,shang2020}. The coexistence of superconductivity and ferromagnetism\cite{Leclair_2005}, together with signatures of broken symmetries, has positioned this system as a candidate platform for exotic superconductivity. Nb on the other hand is an elemental superconductor with dominant conventional s-wave pairing symmetry\cite{ikushima1969,allen1975,laver2009,aceves2024}. 

In this work, we investigate the vortex dynamics and transverse (Hall-like) voltage response in Ni/Bi heterostructures and Nb thin films, with the goal of elucidating the microscopic origin of the transverse signal in superconductors. We show that this transverse voltage is a generic feature of the mixed state, arising from the net force balance on moving vortices — comprising the viscous drag force, the Magnus force, and the Lorentz force exerted by the applied magnetic field on the vortex-induced supercurrents.

Ni/Bi bilayer heterostructures were synthesized by electron-beam evaporation of Ni and using a resistive source for Bi in ultra-high-vacuum with a base pressure of approximately 4 $\times$ 10$^{-8}$ mbar. To suppress interdiffusion and minimize alloy formation at the interface, the films were deposited onto substrates mounted on a liquid-nitrogen-cooled holder, thereby reducing adatom mobility during growth. This low-temperature deposition condition limits atomic diffusion and promotes the formation of a well-defined layered structure. An AlO$_x$ capping layer was deposited in situ to protect the films from surface oxidation and environmental degradation. Nb thin films were sputtered at a base pressure of e-8 mbar. For transport measurements, the samples were patterned into Hall-bar geometries using a laser scribing technique. Conventional photolithographic processing can introduce chemical contamination or thermal damage that degrades superconducting properties; therefore, LASER patterning was employed for Ni/Bi films. Nb thin films were patterned into Hall bar via focused ion beam lithography technique. Electrical contacts were made by bonding the samples to chip carriers using pressed indium providing low-resistance ohmic contacts to the superconducting layers.

\begin{figure}[tbh]
\includegraphics[width=1\columnwidth]{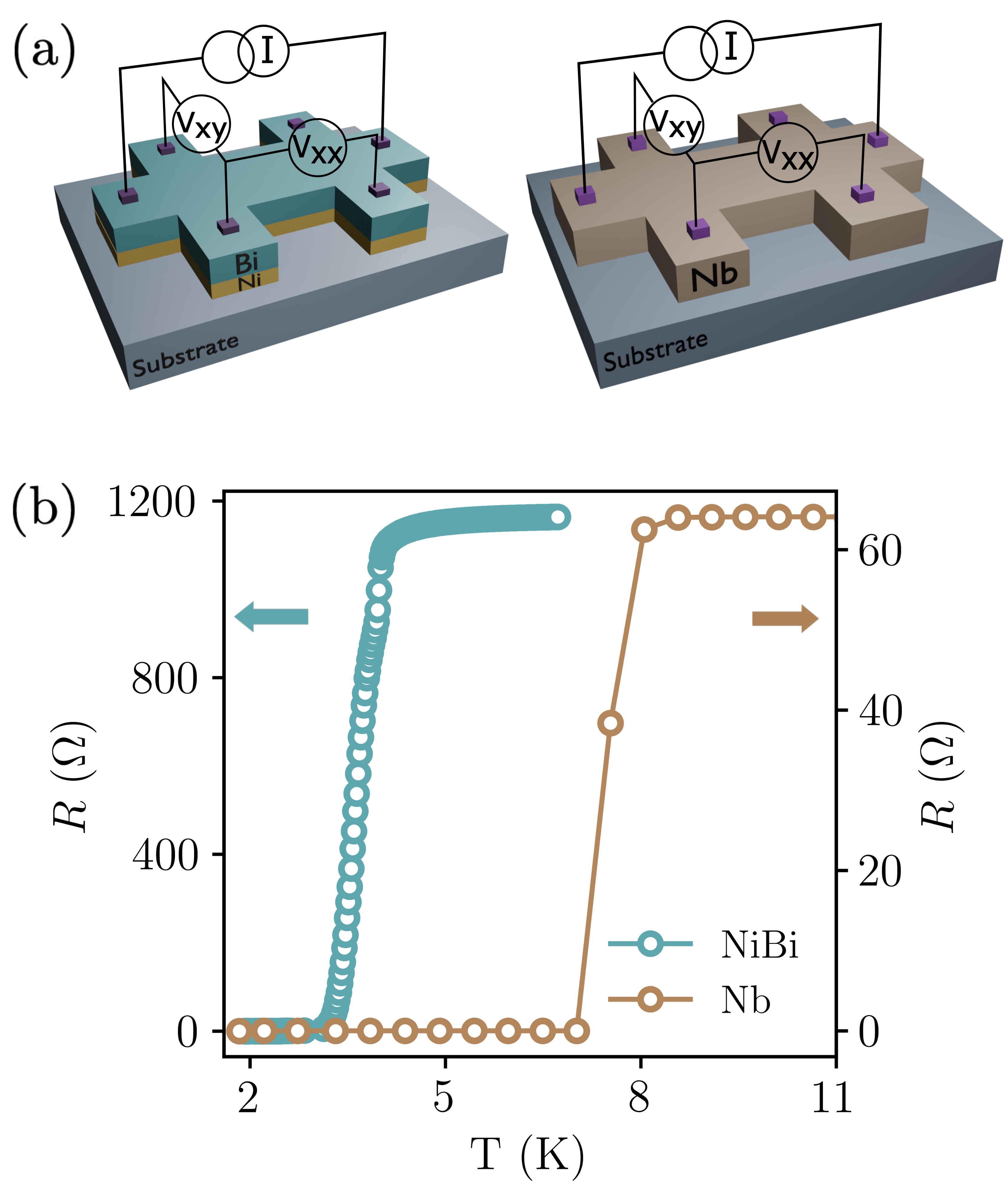} 
\caption{(a) Cartoon schematic of the Hall bar of Ni/Bi and that of Nb, with measurement scheme employed. The width of the Hall bar was $100~\mu m$ distance between voltage probe was $20\mu m$. (b) Resistance versus temperature curves of the two Hall bar devices. }
\label{fig:cartoon}
\end{figure}

Figure~\ref{fig:cartoon}(a) shows the Hall-bar geometry used for transport measurements on the Ni/Bi bilayer and the Nb thin film, with current $I$ applied along the bar and longitudinal ($V_{xx}$) and transverse ($V_{xy}$) voltages measured across dedicated probe pairs. The Ni/Bi bilayer consists of a 2~nm Ni layer grown on a 10~nm Bi layer, patterned into the same device architecture as the Nb (thickness: 20 nm) reference film to allow direct comparison of their transport response under identical measurement conditions. Figure~\ref{fig:cartoon}(b) shows the resistance as a function of temperature for both devices. Each curve shows a sharp drop from a finite normal-state resistance to zero, marking the onset of the superconducting transition; the Ni/Bi bilayer becomes superconducting at a lower temperature ($\approx$ 3.22~K) and with a higher normal-state resistance than the Nb film (transition temperature $T_c~\approx~7.01$ K). 

\begin{figure*}[htbp]
    \centering
    \includegraphics[width=\textwidth]{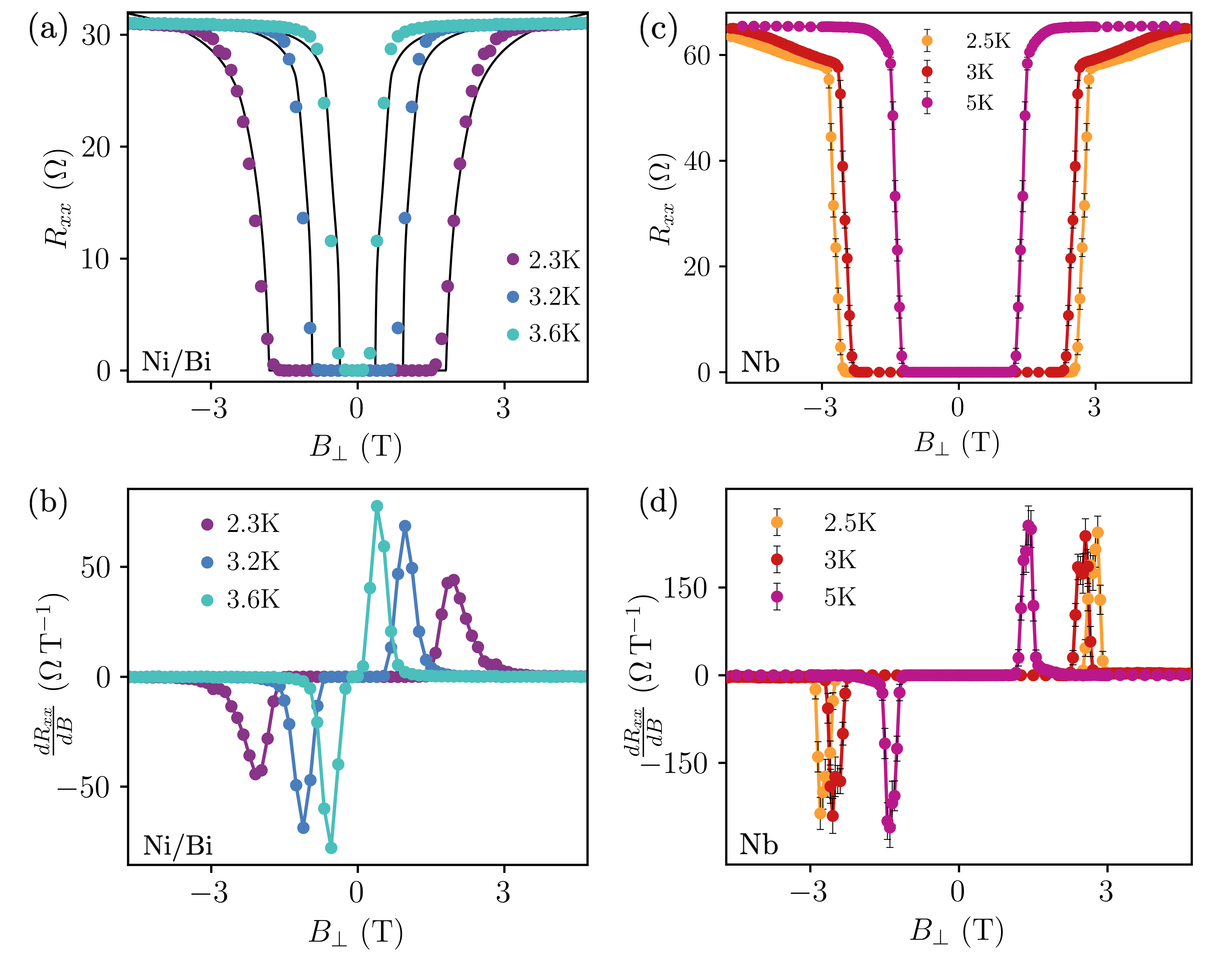}
    \caption{Longitudinal resistance $R_{xx}$ as a function of out-of-plane magnetic field $B_\perp$ for (a) the Ni/Bi bilayer at 2.3~K, 3.2~K, and 3.6~K, and (c) the Nb thin film at 2.5~K, 3~K, and 5~K. In both systems, $R_{xx}$ transitions sharply from the normal-state value to zero as the field is swept through $H_{c2}(T)$, with the transition broadening at temperatures closer to $T_c$ due to thermally assisted vortex depinning. Solid black lines  show the fit with the equation (7). (b,d) Derivative curves corresponding to (a,c) respectively. The noise-level error bars are less than 1\% for (a, b); for fig. (c, d) error bars are shown explicitly.}
    \label{fig:Rxx}
\end{figure*}
Figure~\ref{fig:Rxx} shows the longitudinal resistance $R_{xx}$ as a function of out-of-plane magnetic field $B_\perp$ for the Ni/Bi bilayer [Figure 2 (a)] and the Nb thin film [Figure 2  (c)], measured at several temperatures below their respective zero-field $T_c$. In both devices, $R_{xx}$ exhibits a sharp field-driven transition from the normal-state plateau to zero resistance as $|B_\perp|$ is reduced through the upper critical field $H_{c2}(T)$. The width of this transition, $\Delta$B [defined as the full width at half maximum (FWHM) of the $dR_{xx}/dB$ versus $B$ curve], decreases monotonically with increasing temperature toward $T_c$ [Figure 2 (b), Figure 2 (d)].  At temperatures closer to $T_c$ (3.6 K for Ni/Bi and 5 K for Nb), enhanced thermal activation promotes vortex depinning, reducing the magnetic field required to mobilize vortices and leading to a sharper resistive transition. In contrast, at lower temperatures (2.3 K for Ni/Bi and 2.5 K for Nb), stronger pinning suppresses vortex motion, requiring a higher magnetic field to overcome the pinning potential and resulting in a broader resistive transition. The close qualitative similarity between the two material systems—despite their differing normal-state resistance and $T_c$—indicates that the same underlying dissipation mechanism, namely field-driven vortex entry and flux flow, governs the resistive transition in both Ni/Bi and Nb.

\begin{figure*}[!ht]
    \centering
    \includegraphics[width=\textwidth]{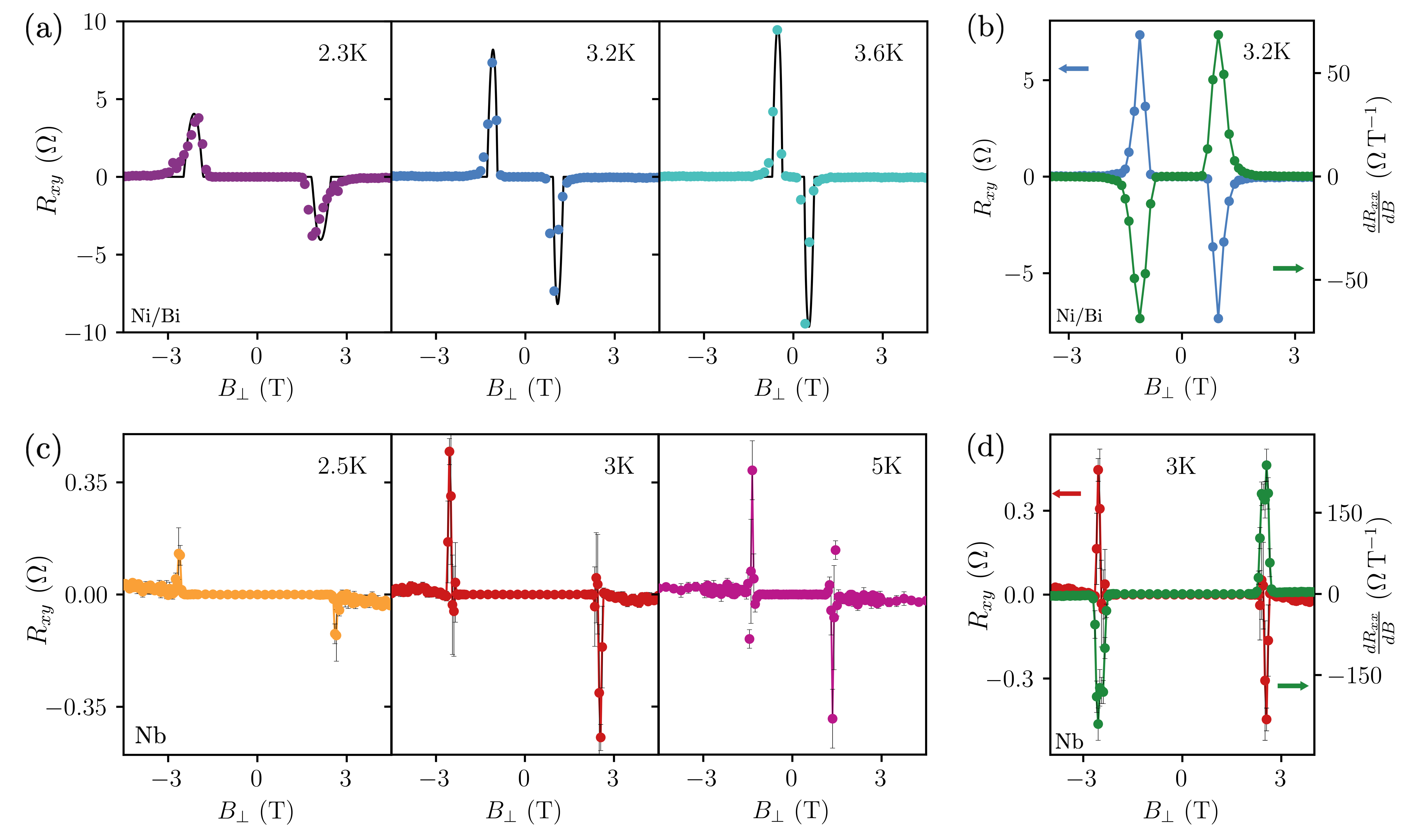}
    \caption{Transverse resistance $R_{xy}$ as a function of out-of-plane magnetic field for (a) the Ni/Bi bilayer and (c) the Nb thin film, at the temperatures indicated in each panel. Solid black lines  show the fit with the equation (8).(b, d) Overlay of $R_{xy}(B)$ (left axis) with the field derivative of the longitudinal resistance, $dR_{xx}/dB$ (right axis), for Ni/Bi at 3.2~K (b) and Nb at 3~K (d). The noise-level of the error bars in (a, b) is lesser than 1\% of the magnitude; for (c, d) the error bars are explicitly indicated in the plots.}
    \label{fig:Rxy}
\end{figure*}

Figure~\ref{fig:Rxy} presents the transverse resistance $R_{xy}$ as a function of $B_\perp$ for Ni/Bi [Fig. 3 (a)] and Nb [Fig. 3(c)], alongside a comparison with the field derivative of the longitudinal resistance, $dR_{xx}/dB$, at a representative temperature for each system [Fig. 3 (b) and (d)]. Two key features emerge: (i) $R_{xy}$ becomes finite for $H > H_{c1}$, and (ii) the peaks are antisymmetric in field, $R_{xy}(H_\perp) = -R_{xy}(-H_\perp)$, consistent with dissipative vortex motion, where moving vortices generate a transverse electric field. In both materials, magnitude of $R_{xy}$ is vanishingly small in the fully superconducting as well as in the normal state, but develops a pronounced, field-antisymmetric peak-and-dip feature within the resistive transition region identified in Fig.~\ref{fig:Rxx}. Figure 3 (b),(d) show that the extrema of $R_{xy}$ coincide closely in field with the extrema of $dR_{xx}/dB$ for both Ni/Bi and Nb, indicating that the transverse signal is largest precisely where the vortex population and vortex mobility are changing most rapidly. This correspondence, observed in both a hetero-structure incorporating a ferromagnetic layer (Ni/Bi) and a conventional elemental superconductor (Nb), demonstrates that the transverse voltage is not an independent transport channel but is directly tied to the same field-dependent vortex dynamics responsible for the longitudinal dissipation. We note that this correspondence between two chemically and structurally unrelated systems, fabricated via independent lithographic routes (laser scribing for Ni/Bi and focused-ion-beam milling for Nb) with independent probe geometries, argues against a common origin in device-specific current-path inhomogeneity \cite{Segal2011,sengupta2025}, which would require two independently fabricated devices to coincidentally exhibit comparable degrees of misalignment reproducing not only the lineshape but its systematic field and temperature evolution. The vortex-flow mechanism, by contrast, is a generic consequence of the mixed state expected in any type-II superconductor near $T_c$, and thus offers the more natural and parsimonious explanation for the shared phenomenology observed here, without excluding a subdominant inhomogeneity contribution superposed on the intrinsic vortex-flow signal.

Figure~4 illustrates the forces acting on a vortex in a type-II superconductor. 
In the absence of an external driving force, vortices can remain pinned at 
defects or inhomogeneities due to a pinning potential. When an external magnetic field and a bias current are applied, a Lorentz force, $\mathbf{F}_{L}$, acts on the vortices. Once the Lorentz force exceeds the pinning force, the vortices become depinned and start moving with a velocity $\mathbf{v}_{L}$ \cite{bardeen1965,blatter1994vortices}.

During their motion, the vortices experience forces arising from their 
interaction with the surrounding environment. One component, 
$\mathbf{F}_{\mathrm{env}}^{\parallel}$, acts antiparallel to the vortex 
velocity and therefore opposes its motion. This component corresponds to the 
viscous drag force. In addition, there can be a transverse environmental force, 
$\mathbf{F}_{\mathrm{env}}^{\perp}$, which acts perpendicular to the vortex 
velocity, analogous to a Magnus-type force. The combined action of these 
longitudinal and transverse forces determines the direction and dynamics of the moving vortex.\cite{sonin1996,ao1997,ao1999}
\begin{figure}[tbh]
\includegraphics[width=1\columnwidth]{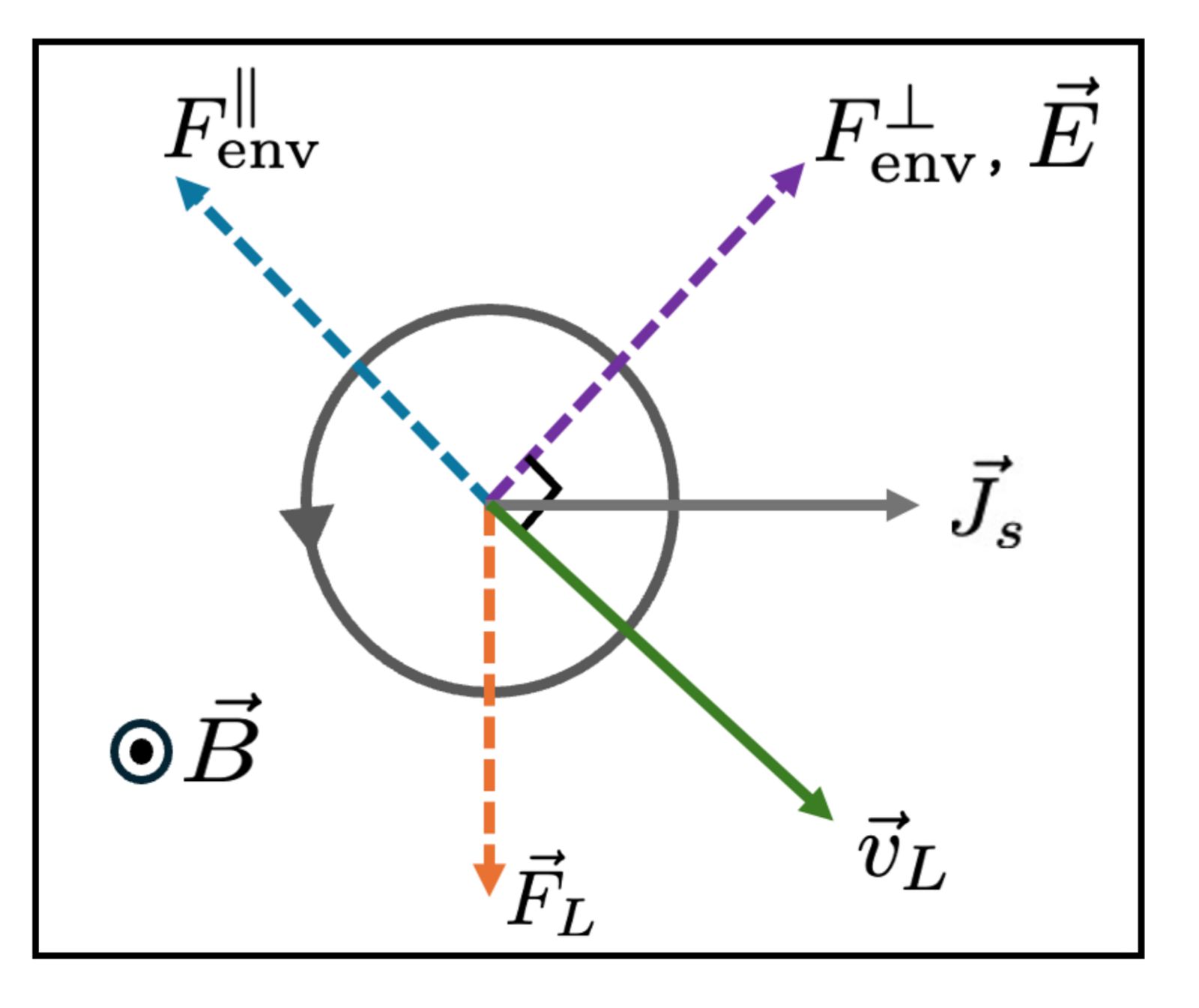} 
\caption{Force diagram representing all the possible forces acting on a vortex.}
\label{fig:device}
\end{figure}

We explain this interpretation in depth within the framework of flux-flow resistivity~\cite{kopnin1995, kopnin2001theory}.  The Lorentz force acting on a vortex is given by
\begin{equation}
\bm{F}_L = \phi_0\, \bm{J} \times \hat{\bm{k}},
\end{equation}
where $\phi_0$ is the flux quantum, $\bm{J}$ is the applied current density, and $\hat{\bm{k}}$ is the unit vector along the magnetic-field direction.

The viscous drag force parallel to the vortex velocity,\cite{bardeen1965}
\begin{equation}
\bm{F}_{\rm env}^{\parallel} = -\eta\, \bm{v}_L,
\end{equation}
where $\eta$ is the viscous drag coefficient associated with quasiparticle dissipation within the vortex core. The transverse reactive force,\cite{ao1997,kopnin2002vortex}
\begin{equation}
\bm{F}_{\rm env}^{\perp} = -\alpha_H\, \bm{v}_L \times \hat{\bm{k}},
\end{equation}
where $\alpha_H$ is the Hall coefficient characterizing the transverse response of the vortex to the applied current. A steady-state flux flow regime is established when the Lorentz force is balanced by the forces exerted by the environment resulting in the equation of motion as below:
\begin{equation}
\eta \bm{v}_L + \alpha_H \left(\bm{v}_L \times \hat{\bm{k}} \right) = \bm{F}_L.
\end{equation}

Solving this equation (by taking a cross product with $\hat{\bm{k}}$ and solving the resulting simultaneous linear equations) yields
\begin{equation}
\bm{v}_L =
\frac{1}{\eta^2 + \alpha_H^2}
\left(
\eta\, \bm{F}_L +
\alpha_H \left(\hat{\bm{k}} \times \bm{F}_L \right)
\right).
\end{equation}
The vortex velocity therefore contains two components: a dissipative component parallel to the Lorentz force and a transverse component arising from the Hall term.

The motion of vortices generates an electric field through the Josephson relation,
\begin{equation}
\bm{E} = \bm{B} \times \bm{v}_L,
\end{equation}
where $\bm{B}$ is the magnetic field. The resulting longitudinal and transverse resistivities are
\begin{align}
R_{xx} &= \frac{\phi_0 \eta B}{\eta^2 + \alpha_H^2}, \\
R_{xy} &= \frac{\phi_0 \alpha_H B}{\eta^2 + \alpha_H^2}.
\end{align}

These expressions follow from a force-balance description of a single vortex and predict a resistivity that is proportional to $B$, consistent with the Bardeen-Stephen picture of flux flow.\cite{bardeen1965} However, the above expressions describe the motion of an isolated vortex and therefore do not fully capture the complexity of vortex dynamics in real superconductors.  In practice, the system with vortices forms a collective ensemble whose behavior is well influenced by vortex–vortex interactions and by pinning due to intrinsic disorder.\cite{blatter1994vortices,blatter1997} The relative importance of these effects evolves with magnetic field as the vortex density changes. It is therefore useful to consider how the physical state of the vortex system evolves across the mixed-state field range.

For magnetic fields just above the lower critical field $H_{c1}$, vortices begin to enter the sample. At these low vortex densities, strong pinning can immobilize a large fraction of vortices, suppressing vortex motion. At intermediate magnetic fields, the vortex density increases and vortex-vortex repulsion becomes significant, effectively reducing the influence of pinning and allowing vortices to move more freely. This regime corresponds to the onset of flux flow. At still higher fields, approaching the upper critical field $H_{c2}$, the superconducting order parameter is progressively suppressed and the vortex cores begin to overlap, causing the transport response to approach that of the normal state.

In order to account for these effects beyond the simple Bardeen-Stephen description, we incorporate vortex-vortex interactions and pinning phenomenologically through a magnetic-field dependence of the viscous drag coefficient $\eta(H)$ and the transverse coefficient $\alpha_H(H)$, given by
\begin{align}
    \eta(H) &= \epsilon_0 H_2\left(1+\epsilon_1\frac{|B|}{H_2}\right), \\
    \alpha(H) &= \alpha_0\frac{B}{H_2}
    \left(1-\frac{|B|}{H_2}\right)
    \theta(H_2-|B|),
\end{align}
where $B=\xi H_2 u\,\theta(u)$, $\xi=\mathrm{sign}(H)$, and
$u=(|H|-H_1)/(H_2-H_1)$. 
The viscosity, $\eta(H)$, is considered to be a linearly increasing function starting from a non-zero value at $H_{c1}$. The transverse response coefficient, $\alpha_H(H)$, is modeled, incorporating Onsager reciprocity, as increasing from zero at $H_{c1}$, reaching a maximum, and decreasing back to zero at $H_{c2}$. Figures~2 (a) and 3(a) show that the model considered here describes the experimentally measured longitudinal and transverse resistance quite well. This approach captures the evolution from the pinned regime at low fields to the flux-flow regime at intermediate fields, and finally to the normal-state response as the system approaches $H_{c2}$. The parameters $H_1$ and $H_2$ correspond to the fields at which $R_{xx}$ begins to increase from zero and approaches saturation, respectively. Their values, estimated from the $R_{xx}$ data, are used only as initial guesses (and to set the bounds) for the fit. All five parameters $\{\alpha_0,\epsilon_0,\epsilon_1,H_1,H_2\}$ are then optimized simultaneously by fitting the model to the experimental data. The resulting values are summarized in Table~\ref{tab:Ni/Bi_fit_parameters}
\begin{table}[h!]
\centering
\setlength{\tabcolsep}{6pt}
\caption{Fitting parameters for Ni/Bi obtained using Eqs.~(7) and (8) at different temperatures. The expressions for $\eta (H)$ and $\alpha (H)$  are given by Eqs.~(9) and (10), respectively.}
\label{tab:Ni/Bi_fit_parameters}
\begin{tabular}{cccccc}
\hline
\textbf{$T$ (K)} & $\boldsymbol{\epsilon_0}$ & $\boldsymbol{\epsilon_1}$ &
$\boldsymbol{\alpha_0}$ & $\boldsymbol{H_1}$ & $\boldsymbol{H_2}$ \\
\hline
2.3 & 0.43 & 2.69 & $-0.85$ & 1.80 & 2.47 \\
3.2 & 0.15 & 4.77 & $-0.85$ & 0.93 & 1.28 \\
3.6 & 0.10 & 5.07 & $-0.80$ & 0.36 & 0.71 \\
\hline
\end{tabular}
\end{table}

  $\alpha_H$ represents the net transverse force on a vortex arising from the competition between the bare Magnus force, due to superfluid momentum circulating around the vortex core, and a compensating transverse contribution from normal quasiparticles bound in the core, which partially cancels the Magnus term. This cancellation is expected to be substantial in the dirty limit relevant to our thin-film samples, so that the resulting flux-flow Hall angle, $\tan\theta_H = R_{xy}/R_{xx}$, should be small compared to unity. From our fits, we extract peak Hall angles of $\tan\theta_H \approx$ 0.3 for Ni/Bi (at T = 3.2~K) and 0.015 for Nb (for T = 3~K), consistent with this expectation and in reasonable agreement with values reported for conventional Nb films \cite{kopnin2001theory}. 
  
We emphasize that this transverse resistance, being a bulk dissipative response governed by 
vortex force balance [Eq.~(4)], is not a symmetry-sensitive probe and should not be taken as 
evidence for or against unconventional pairing in Ni/Bi. This is distinct from the 
even-in-field signal reported by Xu, Li, and Chien in $\beta$-Bi$_2$Pd~\cite{Xu2022}, 
attributed to interfacial chirality; our field-antisymmetric $R_{xy}$, tracking $dR_{xx}/dB$ 
[Fig.~3], instead signals conventional vortex flow. Nb ($s$-wave) and Ni/Bi (proposed 
triplet $p$-wave~\cite{Leclair_2005,Gong_2015,Wang_2017}) show quantitatively comparable behavior 
indicates this channel is governed by common vortex dynamics, independent of pairing symmetry, 
and should be accounted for before invoking exotic mechanisms.  The inhomogeneity scenario of Segal et al. depends on specific disorder patterns and can generate either an even-in-field or an odd-in-field transverse voltage, so the parity of the transverse voltage cannot be predicted a priori for a given sample. Its distinctive strength is the predicted proportionality between the transverse voltage and the field derivative of the longitudinal voltage. Our vortex-dynamics mechanism, which invokes Magnus force, pinning, and vortex-vortex interactions (albeit empirically), is more microscopic and generic. Disorder, manifested through vortex pinning, is nonetheless essential to both scenarios. The equivalent-circuit description of Segal et al. may therefore have a microscopic origin in vortex dynamics. Establishing this connection would reconcile the two scenarios but opens pathways for future research in this direction.

We note that the two films likely host different pinning landscapes, consistent with the 
distinct transition widths and field/temperature dependence of the fitted $\eta(H)$ observed 
for Ni/Bi and Nb (Fig.~2). The Ni/Bi bilayer was deposited at reduced adatom mobility on a 
liquid-nitrogen-cooled substrate and patterned by laser scribing, conditions expected to 
favor a higher density of point-defect and grain-boundary pinning sites and possible 
scribe-edge damage, whereas the Nb film was sputtered and patterned by focused-ion-beam 
lithography, a route typically associated with larger grains and pinning dominated by grain 
boundaries. A full microscopic calculation 
of $\alpha_H$ for these specific material systems would require detailed experiments to understand the nature of pinning potential \cite{Aguirre_2025, Barreto_2026} is beyond the scope of the present 
phenomenological, force-balance analysis.

Our systematic study of longitudinal
and transverse magneto-transport in two distinct
type-II superconducting thin-film systems—a Ni/Bi bilayer and Nb films, the transverse resistance Rxy was found to be negligible within the fully superconducting regime, appearing only in the superconducting transition region as sharp, field antisymmetric peak-and-dip features. We found close qualitative and quantitative (derived from our phenomenological model) correspondence in the behavior of the two systems. Notably, this is despite their markedly different normal state resistances, transition temperatures, and material composition— one the unconventional bilayer coupled to a ferromagnet proposed to host spin-triplet pairing, while the other an elemental s-wave superconductor;  both systems exhibit the same antisymmetric transverse response governed by the similar vortex-flow physics. This agreement seems to indicate that the transverse voltage in Ni/Bi could be explained by the conventional
dynamics of Abrikosov vortices. The role of magnetic proximity effect, broken time-reversal symmetry, or unconventional pairing are not influencing the observed behavior, at least noticeably. These results could provide a vortex-flow framework, a material-independent baseline against which transverse transport anomalies can be assessed. Furthermore, it is apparent that one needs to rule out these conventional explanations before attributing such signals to exotic mechanisms.

DS and NSV thank T. V. Ramakrishan for insightful discussions. LPN thanks DST Nano-PDF for fellowship. DS thanks IISc start-up grant, Ministry of Electronics and Technology, Indian Space Research Organization for funding. Authors duly acknowledge funding from INOXCVA and INOX Airproducts for funding via CSR grants. YH and JSM thank NSF funding (DMR- 2218550). Authors are grateful to micro and nano characterization facility, CeNSE and national nanofabrication facility, CeNSE for facilities usage.

\noindent Data Availability\\
The data that supports the findings of this study are available from the corresponding author upon reasonable request.

\bibliography{Reference_R.bib}

@article{Gong_2015,
  author  = {Gong, X.-X. and Zhou, H.-X. and Xu, P.-C. and Yue, D. and Zhu, K. and Jin, X.-F. and Tian, H. and Zhao, G.-J. and Chen, T.-Y.},
  title   = {Possible p-Wave Superconductivity in Epitaxial Bi/Ni Bilayers},
  journal = {Chin. Phys. Lett.},
  year    = {2015},
  volume  = {32},
  pages   = {067402},
  doi     = {10.1088/0256-307X/32/6/067402}
}

@article{abrikosov1957magnetic,
    author = "Abrikosov, A. A.",
    title = "{On the Magnetic Properties of Superconductors of the Second Group}",
    journal = "Sov. Phys. JETP",
    volume = "5",
    pages = "1174--1182",
    year = "1957"
}

@article{brandt2002vortices,
  title={Vortices in superconductors},
  author={Brandt, Ernst Helmut},
  journal={Phys. C: Superconduct.},
  volume={369},
  number={1-4},
  pages={10--20},
  year={2002},
  publisher={Elsevier},
  doi     = {10.1016/S0921-4534(01)01215-1}

}

@article{huebener1994thermal,
  title={Thermal diffusion of quasiparticles and vortices in high-temperature superconductors},
  author={Huebener, RP and Gross, R and Ri, H-C and Gollnik, F},
  journal={Phys. B: Cond. Matt.},
  volume={197},
  number={1-4},
  pages={588--595},
  year={1994},
  publisher={Elsevier},
  doi     = {10.1016/0921-4526(94)90260-7}
  }

@article{van1995vortex,
  title={Vortex dynamics and the Hall anomaly: A microscopic analysis},
  author={van Otterlo, Anne and Feigel'man, Mikhail and Geshkenbein, Vadim and Blatter, Gianni},
  journal={Phys. Rev. Lett.},
  volume={75},
  number={20},
  pages={3736},
  year={1995},
  publisher={APS},
  doi     = {10.1103/PhysRevLett.75.3736}
  }

@article{davydova2024,
  title={Nonreciprocal superconductivity},
  author={Davydova, Margarita and Geier, Max and Fu, Liang},
  journal={Sci. Adv.},
  volume={10},
  number={48},
  pages={eadr4817},
  year={2024},
  publisher={American Association for the Advancement of Science},
  doi     = {10.1126/sciadv.adr4817}

}

@article{matsuda1995hall,
  title={Hall anomaly in the vortex state of La2-xSrxCuO4},
  author={Matsuda, Y and Nagaoka, T and Suzuki, G and Kumagai, K and Suzuki, Minoru and Machida, M and Sera, M and Hiroi, M and Kobayashi, N},
  journal={Phys. Rev. B},
  volume={52},
  number={22},
  pages={R15749},
  year={1995},
  publisher={APS},
  doi     = {10.1103/PhysRevB.52.R15749}

}

@article{nadeem2023,
  title={The superconducting diode effect},
  author={Nadeem, Muhammad and Fuhrer, Michael S and Wang, Xiaolin},
  journal={Nat. Rev. Phys.},
  volume={5},
  number={10},
  pages={558--577},
  year={2023},
  doi     = {10.1038/s42254-023-00632-w}

}

@article{skocpol1975,
  title={Fluctuations near superconducting phase transitions},
  author={Skocpol, WJ and Tinkham, M},
  journal={Rep. Prog. Phys. },
  volume={38},
  number={9},
  pages={1049--1097},
  year={1975},
  doi     = {10.1088/0034-4885/38/9/001}

}

@article{varlamov2018,
  title={Fluctuation spectroscopy: From Rayleigh-Jeans waves to Abrikosov vortex clusters},
  author={Varlamov, AA and Galda, A and Glatz, A},
  journal={Rev. Mod. Phys.},
  volume={90},
  number={1},
  pages={015009},
  year={2018},
  publisher={APS},
  doi     = {10.1103/RevModPhys.90.015009}

}

@article{lang1995study,
  title={Study of superconducting fluctuations in the high-temperature superconductor YBa2Cu3O7 by magneto-transport measurements},
  author={Lang, WJSM},
  journal={Synth. Met.},
  volume={71},
  number={1-3},
  pages={1555--1558},
  year={1995},
  publisher={Elsevier},
  doi     = {10.1016/0379-6779(94)02949-Y}

}

@article{reimann2015,
  title={Visualizing the morphology of vortex lattice domains in a bulk type-II superconductor},
  author={Reimann, T and Muehlbauer, Sebastian and Schulz, Michael and Betz, B and Kaestner, Anders and Pipich, V and B{\"o}ni, P and Gr{\"u}nzweig, C},
  journal={Nat.  Commun.},
  volume={6},
  number={1},
  pages={8813},
  year={2015},
  publisher={Nature Publishing Group UK London},
  doi     = {10.1038/ncomms9813}

}

@article{liu2020magnetic,
  title={Magnetic properties of superconducting phases NiBi and NiBi3 formed during pulsed laser deposition of Ni-Bi films},
  author={Liu, Liying and Xing, Yutao and Merino, ILC and Franceschini, DF and Sol{\'o}rzano, IG and Baggio-Saitovitch, E},
  journal={J. Magn. Magn. Mater.},
  volume={514},
  pages={167275},
  year={2020},
  publisher={Elsevier},
  doi     = {10.1016/j.jmmm.2020.167275}

}

@article{shang2020,
  title     = {Andreev spectroscopy of the triplet-superconductor state in the Bi/Ni bilayer system},
  author    = {Shang, Xin and Liu, Haiwen and Xia, Ke},
  journal   = {Phys. Rev. B},
  volume    = {101},
  number    = {17},
  pages     = {174514},
  year      = {2020},
  publisher = {APS},
  doi       = {10.1103/PhysRevB.101.174514}
}

@article{allen1975,
  title={Transition temperature of strong-coupled superconductors reanalyzed},
  author={Allen, Ph B and Dynes, RC},
  journal={Phys. Rev. B},
  volume={12},
  number={3},
  pages={905},
  year={1975},
  publisher={APS},
  doi     = {10.1103/PhysRevB.12.905}

}

@article{laver2009,
  title={Structure and degeneracy of vortex lattice domains in pure superconducting niobium: A small-angle neutron scattering study},
  author={Laver, Mark and Bowell, CJ and Forgan, EM and Abrahamsen, Asger Bech and Fort, David and Dewhurst, CD and M{\"u}hlbauer, S and Christen, David K and Kohlbrecher, J and Cubitt, R and others},
  journal={Phys. Rev. B},
  volume={79},
  number={1},
  pages={014518},
  year={2009},
  publisher={APS},
  doi     = {10.1103/PhysRevB.79.014518}

}

@article{ikushima1969,
  title={Superconductivity in niobium and niobium-tantalum alloys},
  author={Ikushima, Akira and Mizusaki, Takao},
  journal={J. Phys. Chem. Solids},
  volume={30},
  number={4},
  pages={873--879},
  year={1969},
  publisher={Elsevier},
  doi     = {10.1016/0022-3697(69)90283-2}

}

@article{aceves2024,
  title={Superconductivity in Nb: impact of temperature, dimensionality and cooper-pairing},
  author={Aceves Rodriguez, Uriel Allan and Guimar{\~a}es, Filipe Souza Mendes and Lounis, Samir},
  journal={Nanomaterials},
  volume={14},
  number={3},
  pages={254},
  year={2024},
  publisher={MDPI},
  doi     = {10.3390/nano14030254}

}

@article{sengupta2025,
  title={Transverse resistance due to electronic inhomogeneities in superconductors},
  author={Sengupta, Shamashis and Farhadizadeh, Alireza and Youssef, Joe and Loucif, Sara and Pallier, Florian and Dumoulin, Louis and Saha, Kasturi and Pujari, Sumiran and {\'O}den, Magnus and Marrache-Kikuchi, Claire and others},
  journal={Phys. Rev. B},
  volume={112},
  number={1},
  pages={014514},
  year={2025},
  publisher={APS},
  doi     = {10.1103/bgbf-b39l}

}

@article{vinokur1993,
  title={Scaling of the Hall resistivity in high-T c superconductors},
  author={Vinokur, Valerii M and Geshkenbein, Vadim B and Feigel’man, Mikhail V and Blatter, Gianni},
  journal={Phys. Rev. Lett.},
  volume={71},
  number={8},
  pages={1242},
  year={1993},
  publisher={APS},
  doi     = {10.1103/PhysRevLett.71.1242}

}

@article{ao1999,
  title={Microscopic theory of vortex dynamics in homogeneous superconductors},
  author={Ao, Ping and Zhu, X-M},
  journal={Phys. Rev. B},
  volume={60},
  number={9},
  pages={6850},
  year={1999},
  publisher={APS},
  doi     = {10.1103/PhysRevB.60.6850}

}

@article{ao1997,
  title={Vortex dynamics within the BCS theory},
  author={Ao, Ping and Zhu, X-M},
  journal={Phys. C: Superconduct.},
  volume={282},
  pages={367--370},
  year={1997},
  publisher={Elsevier},
  doi     = {10.1016/S0921-4534(97)00272-4}

}

@article{sonin1996,
  title={The Magnus force and vortex dynamics in superfluids and superconductors},
  author={Sonin, EB},
  journal={Czech. J. Phys.},
  volume={46},
  number={Suppl 2},
  pages={911--912},
  year={1996},
  publisher={Springer},
  doi     = {10.1007/BF02583763}

}

@article{kopnin2002vortex,
  title={Vortex dynamics and mutual friction in superconductors and Fermi superfluids},
  author={Kopnin, Nikolai B},
  journal={Rep. Prog. Phys. },
  volume={65},
  number={11},
  pages={1633--1678},
  year={2002},
  doi     = {10.1088/0034-4885/65/11/202}

}

@article{blatter1997,
  title={Vortex matter},
  author={Blatter, G},
  journal={Phys. C: Superconduc.},
  volume={282},
  pages={19--26},
  year={1997},
  publisher={Elsevier},
  doi     = {10.1016/S0921-4534(97)00197-4}

}

@article{Leclair_2005,
  author  = {LeClair, P. and Moodera, J. S. and Philip, J. and Heiman, D.},
  title   = {Coexistence of Ferromagnetism and Superconductivity in Ni/Bi Bilayers},
  journal = {Phys. Rev. Lett.},
  year    = {2005},
  volume  = {94},
  pages   = {037006},
  doi     = {10.1103/PhysRevLett.94.037006}
}

@article{Wang_2017,
  author  = {Wang, J. and Gong, X.-X. and Yang, G. and Lyu, Z. and Pang, Y. and Liu, G. and Ji, Z. and Fan, J. and Jing, X. and Yang, C. and Qu, F. and Jin, X.-F. and Lu, L.},
  title   = {Anomalous Magnetic Moments as Evidence of Chiral Superconductivity in a Bi/Ni Bilayer},
  journal = {Phys. Rev. B},
  year    = {2017},
  volume  = {96},
  pages   = {054519},
  doi     = {10.1103/PhysRevB.96.054519}
}

@article{Segal2011,
  author  = {Segal, A. and Karpovski, M. and Gerber, A.},
  title   = {Inhomogeneity and transverse voltage in superconductors},
  journal = {Phys. Rev. B},
  volume  = {83},
  pages   = {094531},
  year    = {2011},
  doi     = {10.1103/PhysRevB.83.094531}
}

@article{Xu2022,
  author  = {Xu, Xiaoying and Li, Yufan and Chien, C. L.},
  title   = {Anomalous transverse resistance in the topological superconductor $\beta$-Bi$_2$Pd},
  journal = {Nat. Commun.},
  volume  = {13},
  pages   = {5321},
  year    = {2022},
  doi     = {10.1038/s41467-022-32877-x}
}

@book{kopnin2001theory,
  author    = {Kopnin, N. B.},
  title     = {Theory of Nonequilibrium Superconductivity},
  publisher = {Oxford University Press},
  year      = {2001},
  isbn      = {9780198507888},
  doi       ={10.1093/acprof:oso/9780198507888.001.0001}
}

@article{blatter1994vortices,
  author  = {Blatter, G. and Feigel'man, M. V. and Geshkenbein, V. B. and Larkin, A. I. and Vinokur, V. M.},
  title   = {Vortices in High-Temperature Superconductors},
  journal = {Rev. Mod. Phys.},
  year    = {1994},
  volume  = {66},
  pages   = {1125},
  doi     = {10.1103/RevModPhys.66.1125}
}

@article{wang1991anomalous,
  author  = {Wang, Z. D. and Ting, C. S.},
  title   = {Anomalous Hall Effect Associated with Pinning in High-$\kappa$ Superconductors},
  journal = {Phys. Rev. Lett.},
  year    = {1991},
  volume  = {67},
  pages   = {3618},
  doi     = {10.1103/PhysRevLett.67.3618}
}

@article{bardeen1965,
  author  = {Bardeen, J. and Stephen, M. J.},
  title   = {Theory of the Motion of Vortices in Superconductors},
  journal = {Phys. Rev.},
  year    = {1965},
  volume  = {140},
  pages   = {A1197--A1207},
  doi     = {10.1103/PhysRev.140.A1197}
}

@article{Nozieres1966,
  author  = {Nozi{\`e}res, P. and Vinen, W. F.},
  title   = {The motion of flux lines in type II superconductors},
  journal = {Philosophical Magazine},
  volume  = {14},
  number  = {130},
  pages   = {667--688},
  year    = {1966},
  doi     = {10.1080/14786436608211964}
}

@article{kopnin1995,
  author  = {Kopnin, N. B.},
  title   = {Flux-Flow Hall Effect in Clean Type-II Superconductors},
  journal = {Phys. Rev. B},
  year    = {1995},
  volume  = {51},
  pages   = {15291--15304},
  doi     = {10.1103/PhysRevB.51.15291}
}

@article{Barreto_2026,
  author  = {Barreto, Z. V. and Aguirre, C. A. and Barba-Ortega, J.},
  title   = {Systematic study of the magnetic response of a superconducting meso-bridge under a geometry variation},
  journal = {Phys. Scr.},
  volume  = {101},
  number  = {23},
  pages   = {235906},
  year    = {2026},
  doi     = {10.1088/1402-4896/ae7846},
  publisher = {IOP Publishing}
}

@article{Aguirre_2025,
  author = {Aguirre, C. A. and Faúndez, J. and Díaz, P. and Laroze, D. and Mosquera, A. S. and Costa, N. C. and Barba-Ortega, J.},
  title = {Controlling vortex matter via a superconducting nano-bridge sample},
  journal = {Appl. Phys. A},
  year = {2025},
  volume = {131},
  number = {6},
  pages = {501},
  doi = {10.1007/s00339-025-08644-2},
  
  isbn = {1432-0630}
}

@article{Brandt1995,
  author        = {Brandt, Ernst Helmut},
  title         = {The Flux-Line Lattice in Superconductors},
  journal       = {Rep. Prog. Phys. },
  volume        = {58},
  number        = {11},
  pages         = {1465--1594},
  year          = {1995},
  doi = {10.1088/0034-4885/58/11/003}
}

@article{Pinto2018,
  author    = {Pinto, Nicola and Rezvani, S. Javad and Perali, Andrea and
               Flammia, Luca and Milo{\v{s}}evi{\'c}, Milorad V. and
               Fretto, Matteo and Cassiago, Cristina and De Leo, Natascia},
  title     = {Dimensional crossover and incipient quantum size effects in
               superconducting niobium nanofilms},
  journal   = {Scientific Reports},
  volume    = {8},
  pages     = {4710},
  year      = {2018},
  doi       = {10.1038/s41598-018-22983-6}
}

@article{HarperTinkham1968,
  author    = {Harper, F. E. and Tinkham, M.},
  title     = {The Mixed State in Superconducting Thin Films},
  journal   = {Physical Review},
  volume    = {172},
  pages     = {441--450},
  year      = {1968},
  doi       = {10.1103/PhysRev.172.441}
}

@book{Tinkham2004,
  author    = {Tinkham, M.},
  title     = {Introduction to Superconductivity},
  edition   = {2nd},
  publisher = {Dover Publications},
  address   = {Mineola, NY},
  year      = {2004}
}

\end{document}


\title{Supporting Information for the Article:\\
Vortex Dynamics Driven Transverse Transport in Superconducting Thin Films}

\author{Laxmipriya Nanda}
\affiliation{Centre for Nanoscience and Engineering,
Indian Institute of Science, Bengaluru, Karnataka 560012, India}

\author{Sohini Guin}
\affiliation{Centre for Nanoscience and Engineering,
Indian Institute of Science, Bengaluru, Karnataka 560012, India}

\author{Yasen Hou}
\affiliation{Plasma Science and Fusion Center,
Massachusetts Institute of Technology, Cambridge 02139, USA}

\author{Rajib Sarkar}
\affiliation{Centre for Nanoscience and Engineering,
Indian Institute of Science, Bengaluru, Karnataka 560012, India}

\author{Naresh Shyaga}
\affiliation{Centre for Nanoscience and Engineering,
Indian Institute of Science, Bengaluru, Karnataka 560012, India}

\author{Souvik Banerjee}
\affiliation{Jawaharlal Nehru Centre for Advanced Scientific Research,
Jakkur, Bengaluru}

\author{A. Sundaresan}
\affiliation{Jawaharlal Nehru Centre for Advanced Scientific Research,
Jakkur, Bengaluru}

\author{N. S. Vidhyadhiraja}
\affiliation{Jawaharlal Nehru Centre for Advanced Scientific Research,
Jakkur, Bengaluru}

\author{Jagadeesh S. Moodera}
\affiliation{Plasma Science and Fusion Center,
Massachusetts Institute of Technology, Cambridge 02139, USA}
\affiliation{Department of Physics,
Massachusetts Institute of Technology, Cambridge 02139, USA}

\author{Dhavala Suri}
\email{dsuri@iisc.ac.in}
\affiliation{Centre for Nanoscience and Engineering,
Indian Institute of Science, Bengaluru, Karnataka 560012, India}

\maketitle

\section{Fitting of $R_{xx}$ vs.~$B$ and $R_{xy}$ vs.~$B$
for Nb at 3 K Using Eqs.~(7) and (8) from the Main Text}

\begin{figure}[!ht]
\centering
\includegraphics[width=\textwidth]{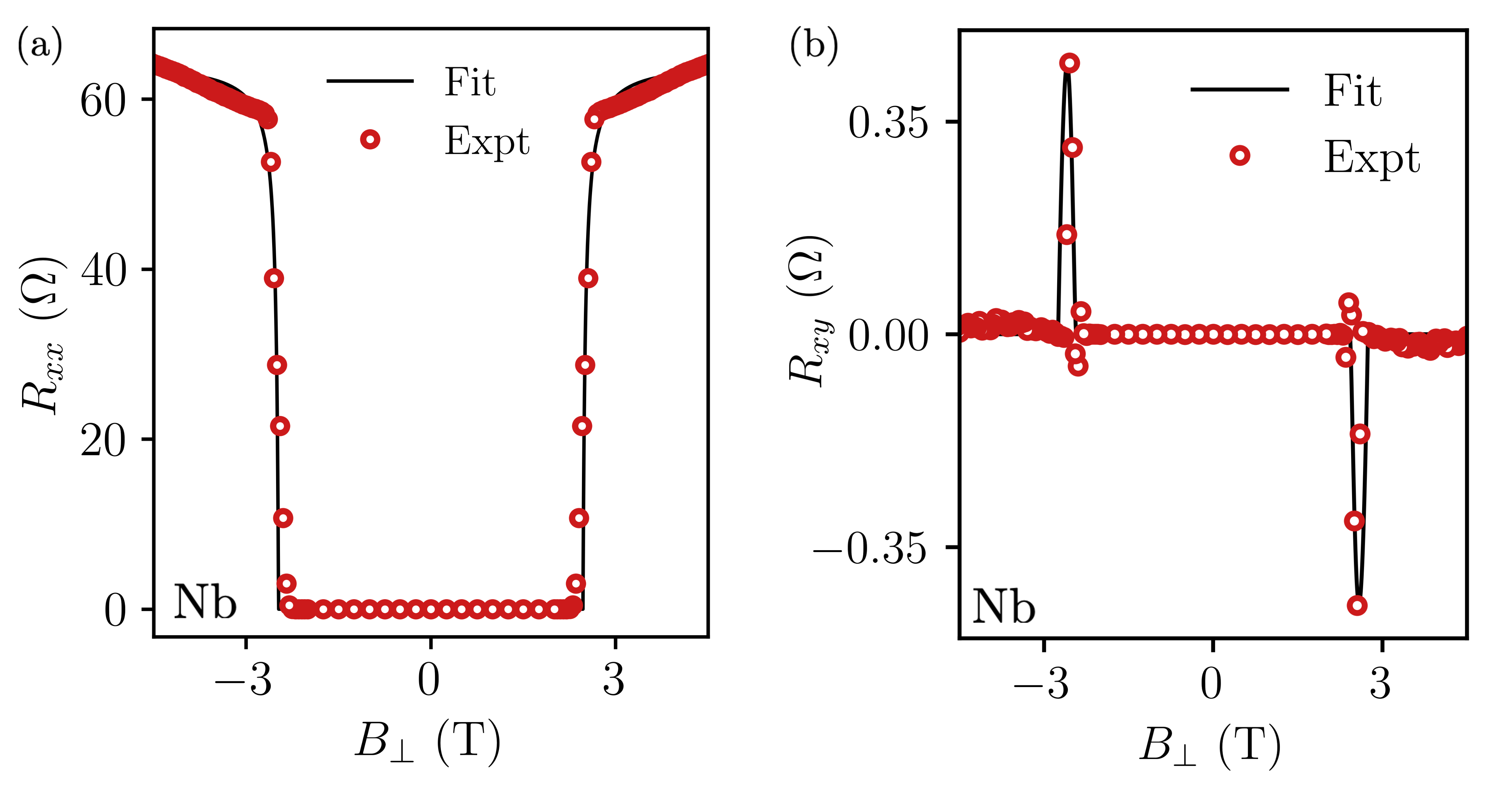}
\caption{
(a) $R_{xx}$ as a function of $B$ for the Nb Hall bar at
$3\,\mathrm{K}$. The solid black line represents the fit to the
experimental data using Eq.~(7) in the main text.
(b) $R_{xy}$ as a function of $B$ for the Nb Hall bar at
$3\,\mathrm{K}$. The solid black line represents the fit to the
experimental data using Eq.~(8) in the main text.
}
\label{fig:SI_Nb}
\end{figure}

\begin{table}[!ht]
\caption{Fitting parameters for Nb using Eqs. (7) and (8) at 3 K .}
\begin{tabular}[t]{|l|l|l|l|l|l|}
\hline $T$ (K) & $\epsilon_0$ & $\epsilon_1$ & $\alpha$ & $H_1$ & $H_2$ \\
\hline 3 & 0.15 & 1.87 & -0.02 & 2.42 & 2.75 \\
\hline
\end{tabular}
\end{table}

\clearpage

\section{In-plane Field Measurements}

We also explored the response of the transverse voltage when
the magnetic field was applied in the plane of the sample,
with the current and magnetic field perpendicular to each other.
The temperature was set close to $T_c$ to access the
superconductor--normal-state transition.

At $T = 6.1\,\mathrm{K}$, a similar transverse peak was observed
under an in-plane magnetic field. We attribute this observation
to possible misalignment during sample mounting. Even a
misalignment of $1$--$2^\circ$ can introduce a perpendicular
field component, allowing vortices to experience forces
associated with vortex motion, including Lorentz and
Magnus-type forces.

\begin{figure}[!ht]
\centering
\includegraphics[width=0.6\textwidth]{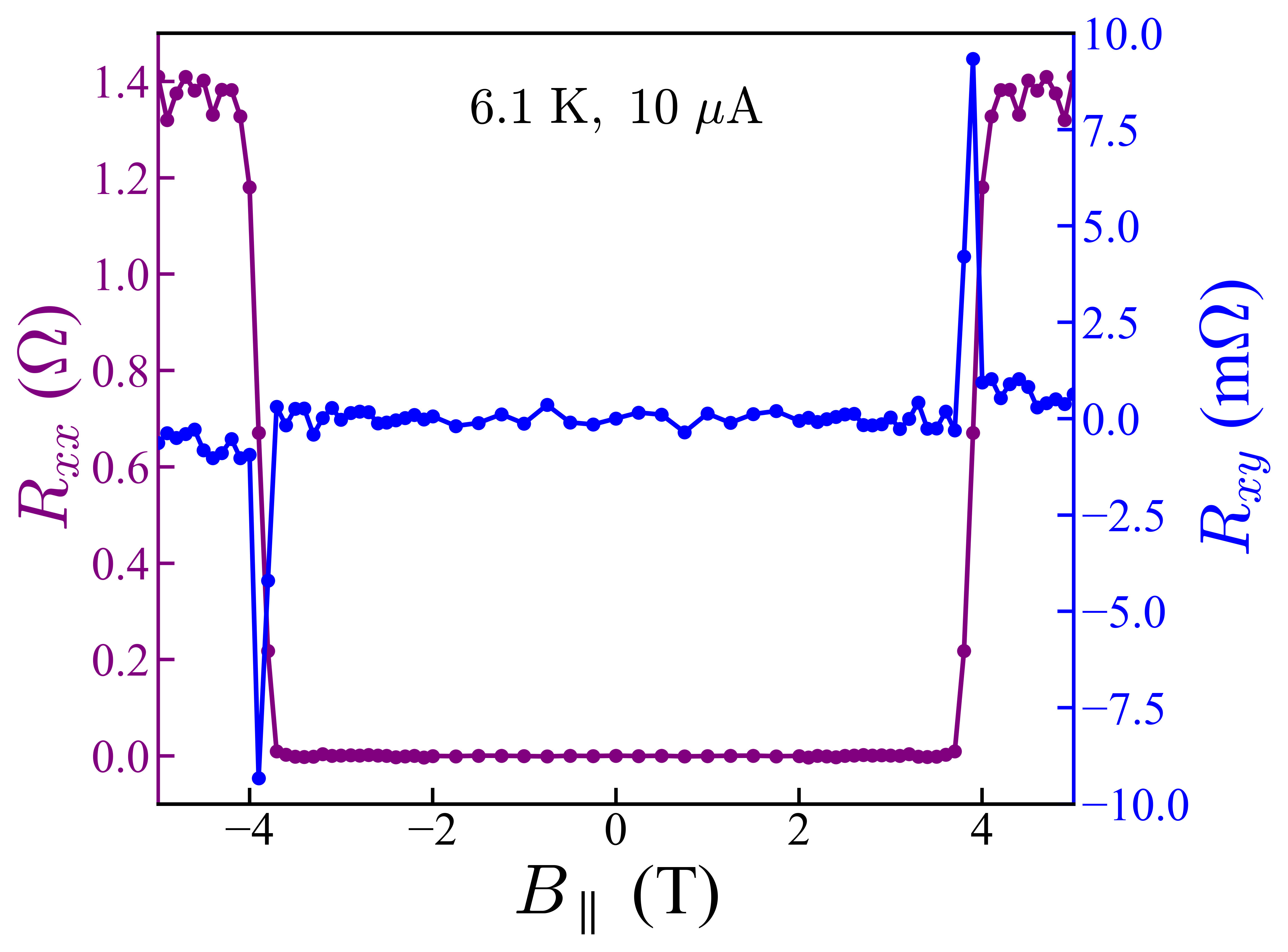}
\caption{
Comparison of $R_{xx}$ (left $y$-axis) as a function of $B$
and $R_{xy}$ (right $y$-axis) as a function of $B$ for the
Nb Hall bar at $6.1\,\mathrm{K}$.
}
\label{fig:SI_Nb_6.1K}
\end{figure}

For sputtered Nb films of comparable thickness
($d \approx 20\,\mathrm{nm}$), the Ginzburg--Landau coherence
length obtained from $H_{c2\perp}(T)$ is
$\xi(0) \simeq 8\,\mathrm{nm}$, reduced from the bulk value.
Such films display a 3D-to-2D crossover below
$d \approx 25\,\mathrm{nm}$ \cite{Pinto2018}.

Scaling as
\[
\xi(T) = \frac{\xi(0)}{\sqrt{1-T/T_c}}
\]
gives $\xi(6.1\,\mathrm{K}) \approx 22\,\mathrm{nm} \approx d$ and
\[
\mu_0 H_{c2\perp}(6.1\,\mathrm{K}) = \frac{\phi_0}{2\pi\xi^2} \approx 0.7\,\mathrm{T}.
\]

The film is therefore in the thin-film limit, where
\[
H_{c2\parallel} = \frac{\sqrt{12}\,\phi_0}{2\pi\xi d}
\]
\cite{Tinkham2004,HarperTinkham1968}.

We also note that current redistribution and inhomogeneity,
as suggested by Refs.~\cite{Segal2011,sengupta2025},
cannot be entirely ruled out. However, we consider a
vortex-related mechanism the most natural explanation
for our observations.

\clearpage

\section{Comparison with Other Similar Works}

\begin{table}[ht!]
\centering
\small
\caption{Comparison of the present work with Segal \textit{et al.} and Sengupta \textit{et al.}.}
\label{tab:comparison}

\renewcommand{\arraystretch}{1.2}

\resizebox{\textwidth}{!}{%
\begin{tabular}{|p{2.8cm}|p{4.2cm}|p{4.2cm}|p{4.2cm}|}
\hline
& \textbf{Segal \textit{et al.} (PRB 2011)}
& \textbf{Sengupta \textit{et al.} (PRB 2025)}
& \textbf{This work} \\
\hline

Materials
& Pb, YBaCuO, granular NiPb
& Nb, NbN (conventional)
& Ni/Bi bilayer (proposed triplet, with ferromagnet), Nb \\
\hline

Field parity of signal
& Even (plus zero-field signal vs.\ $T$)
& Even
& Odd (antisymmetric) \\
\hline

Proposed origin
& Structural inhomogeneity: diagonal spread of $T_c/H_{c2}$ bends current paths
& Emergent electronic inhomogeneity near vortex-lattice melting, not tied to structure
& Flux-flow Hall: force balance of Lorentz, viscous drag, and Magnus-type forces \\
\hline

Sign behavior
& Set by disorder geometry; reversed with engineered thickness gradient
& Changes with temperature and between samples
& Same sign relation across temperatures and both materials \\
\hline

Role of vortices
& Explicitly excluded (same signal with Lorentz-force-free in-plane field)
& Vortex melting invoked for the emergent texture, not for Hall-type flux flow
& Central mechanism \\
\hline

Model
& Four-resistor network plus $100\times300$ resistor grid, one fit parameter
& Qualitative two-fluid/GL argument
& Single-vortex force balance with phenomenological $\eta(H)$ and $\alpha_H(H)$ \\
\hline

Sample scale
& 200 nm films
& Nb: 55 nm, $w=40\,\mu\mathrm{m}$; NbN: 205 nm, mm-scale contacts
& Ni/Bi: 2 nm/10 nm; Nb: 20 nm; $w=100\,\mu\mathrm{m}$ \\
\hline
\end{tabular}%
}
\end{table}

\clearpage

\bibliographystyle{apsrev4-1}
\bibliography{Reference_R}